\documentclass[prb, reprint, superscriptaddress, showpacs]{revtex4-1}
\usepackage[dvips]{graphicx}
\usepackage{color}
\usepackage{amsmath}
\usepackage[breaklinks]{hyperref}
\usepackage[all]{hypcap}

\begin{document}

\title{Nuclear magnetic resonance study of the magnetic-field-induced \\ ordered phase in the NiCl$_{2}$-4SC(NH$_{2}$)$_{2}$ compound}

\author{R\'{e}mi Blinder}
\altaffiliation[Present address: ]{CEA, INAC, SPRAM (UMR5819 CEA/CNRS/UJF), 17 rue des Martyrs, 38054 Grenoble cedex 9, France}
\affiliation{Laboratoire National des Champs Magn\'etiques Intenses, LNCMI-CNRS (UPR3228), \\ EMFL, UGA, UPS, and INSA, Bo\^{i}te Postale 166, 38042, Grenoble Cedex 9, France}

\author{Maxime Dupont}
\affiliation{Laboratoire de Physique Th\'{e}orique, IRSAMC, Universit\'{e} de Toulouse, CNRS, 31062 Toulouse, France}

\author{Sutirtha Mukhopadhyay}
\altaffiliation{Present address: BAT-SOL Equipment and Technology, Vashi, Navi Mumbai - 400 705, India}
\affiliation{Laboratoire National des Champs Magn\'etiques Intenses, LNCMI-CNRS (UPR3228), \\ EMFL, UGA, UPS, and INSA, Bo\^{i}te Postale 166, 38042, Grenoble Cedex 9, France}

\author{Mihael S. Grbi\'{c}}
\affiliation{Department of Physics, Faculty of Science, University of Zagreb, Bijeni\v {c}ka 32, Zagreb HR 10000, Croatia}
\affiliation{Laboratoire National des Champs Magn\'etiques Intenses, LNCMI-CNRS (UPR3228), \\ EMFL, UGA, UPS, and INSA, Bo\^{i}te Postale 166, 38042, Grenoble Cedex 9, France}

\author{Nicolas Laflorencie}
\affiliation{Laboratoire de Physique Th\'{e}orique, IRSAMC, Universit\'{e} de Toulouse, CNRS, 31062 Toulouse, France}

\author{Sylvain Capponi}
\affiliation{Laboratoire de Physique Th\'{e}orique, IRSAMC, Universit\'{e} de Toulouse, CNRS, 31062 Toulouse, France}

\author{Hadrien Mayaffre}
\affiliation{Laboratoire National des Champs Magn\'etiques Intenses, LNCMI-CNRS (UPR3228), \\ EMFL, UGA, UPS, and INSA, Bo\^{i}te Postale 166, 38042, Grenoble Cedex 9, France}

\author{Claude Berthier}
\affiliation{Laboratoire National des Champs Magn\'etiques Intenses, LNCMI-CNRS (UPR3228), \\ EMFL, UGA, UPS, and INSA, Bo\^{i}te Postale 166, 38042, Grenoble Cedex 9, France}

\author{Armando Paduan-Filho}
\affiliation{Instituto de F\'isica, Universidade de S{\~a}o Paulo, 05315-970 S{\~a}o Paulo, Brazil}

\author{Mladen Horvati\'c}
\email{mladen.horvatic@lncmi.cnrs.fr}
\affiliation{Laboratoire National des Champs Magn\'etiques Intenses, LNCMI-CNRS (UPR3228), \\ EMFL, UGA, UPS, and INSA, Bo\^{i}te Postale 166, 38042, Grenoble Cedex 9, France}

\begin{abstract} Nuclear magnetic resonance (NMR) study of the high magnetic field ($H$) part of the Bose-Einstein condensed (BEC) phase of the quasi-onedimensional (quasi-1D) antiferromagnetic quantum spin-chain compound NiCl$_2$-4SC(NH$_2$)$_2$ (DTN) was performed. We precisely determined the phase boundary, $T_\textrm{c}(H)$, down to 40 mK; the critical boson density, $n_\textrm{c}(T_\textrm{c})$; and the absolute value of the BEC order parameter $S_\perp$ at very low temperature ($T$~=~0.12~K). All results are accurately reproduced by numerical quantum Monte Carlo simulations of a realistic three-dimensional (3D) model Hamiltonian. Approximate analytical predictions based on the 1D Tomonaga-Luttinger liquid description are found to be precise for $T_\textrm{c}(H)$, but less so for $S_\perp(H)$, which is more sensitive to the strength of 3D couplings, in particular close to the critical field. A mean-field treatment, based on the Hartree-Fock-Popov description, is found to be valid only up to $n_\mathrm{c} \cong 4$\% ($T < 0.3$~K), while for higher $n_\textrm{c}$ boson interactions appear to modify the density of states.   \end{abstract}

\date{\today}





\maketitle

Quantum phase transitions, i.e., phase transitions that are driven by quantum, rather than thermal, fluctuations, are one of the topical subjects in condensed matter physics.\cite{sachdevquantumphasetransition, coleman2005, sachdevkeimer, si2010} There are numerous experimental investigations of such transitions as a function of an external control parameter, such as magnetic field ($H$), pressure, or chemical composition. The NiCl$_2$-4SC(NH$_2$)$_2$ (DTN) quantum magnet has long been studied in this respect.\cite{filho1981, filho2004, zapf2006, zvyagin2007, mukhopadhyay2012, tsyrulin2013, zheludev2015} The system consists of weakly coupled chains of $S=1$ spins, borne by Ni$^{2+}$ ions, subject to the Hamiltonian\cite{zvyagin2007}
\begin{equation}
 \mathcal{H}=\sum\limits_{\boldsymbol{r}} \hspace{2pt} \Big\{   \sum\limits_{\boldsymbol{v}=\boldsymbol{a},\boldsymbol{b},\boldsymbol{c}} \hspace{-5pt} J_{\boldsymbol{v}}\boldsymbol{\hat{S}}_{\boldsymbol{r}}\cdot\boldsymbol{\hat{S}}_{\boldsymbol{r}+\boldsymbol{v}} \Big\}  + D (\hat{S}^{z}_{\boldsymbol{r}})^2 - g\mu_\mathrm{B} H \hat{S}^{z}_{\boldsymbol{r}}, \label{Hamiltonian}
\end{equation}
where summations are performed over all lattice positions ($\boldsymbol{r}$) and unit cell vectors ($\boldsymbol{v}$). Equation~(\ref{Hamiltonian}) shows that the spins are subject to an easy-plane anisotropy [the $D (\hat{S}^{z}_{\boldsymbol{r}})^2$ term, where  $D/k_B~=~8.9$~K] and the nearest-neighbor Heisenberg interaction \mbox{($J_{\boldsymbol{v}}\boldsymbol{\hat{S}}_{\boldsymbol{r}}\cdot\boldsymbol{\hat{S}}_{\boldsymbol{r}+\boldsymbol{v}}$),} preferentially along the chain ($c$-axis direction), $J_{c}/k_B =2.2$~K. Also, an interchain coupling $J_{ab}/k_\textrm{B} = 0.18$~K is present, which, because of the tetragonal symmetry, is equivalent for the $a$ and $b$ directions.  As the antiferromagnetic  (AF) couplings differ by an order of magnitude ($J_{c}/J_{ab}\cong 12$), the system can be considered as quasi-onedimensional (quasi-1D). Between critical fields $H_\textrm{c1} =2.1$~T and $H_\textrm{c2} =12.32$~T,\cite{filho2004,zapf2006,mukhopadhyay2012} and at low temperature ($T$) it displays a three-dimensional (3D) AF ordered phase that can be described as a Bose-Einstein condensation (BEC) of the spin degrees of freedom (belonging to the 3D \textit{XY} universality class).\cite{naturephys2008} In the BEC phase a transverse AF magnetic moment develops, corresponding to the BEC order parameter.\cite{affleck1991, nikuni2000, naturephys2008, zapf2014} Although the existence of such a state is intrinsically a 3D phenomenon, a pronounced 1D character is known to affect some of its properties in a nontrivial and interesting way.\cite{giamarchi1999, bouillot2011} The spin ladders CuBr$_4$(C$_5$H$_{12}$N)$_2$ (BPCB)~\cite{klanjsek2008, bouillot2011} and (C$_7$H$_{10}$N$_2$)CuBr$_4$ (DIMPY),\cite{hong2010prl, jeong2013, jeong2016} whose anisotropy of coupling constants (1D character) is $\sim$10 times stronger than in DTN, were successfully described using the 1D Tomonaga-Luttinger liquid (TLL) framework,\cite{giamarchiqphys1D_book} enhanced by considering the transverse couplings,\cite{giamarchi1999, klanjsek2008, bouillot2011} which induce a long-range order at finite temperature. We recall here that both Hamiltonians of a spin-1 chain in the $D \gg J_c$ limit and a spin-1/2 ladder in the $J_\textrm{rung} \gg J_\textrm{leg}$ limit can be reduced, for low temperature, to the same basic \textit{XXZ} spin-1/2 chain model and are thus equivalent -- as much as the strong $D$ and $J_\textrm{rung}$ limits are respected.\cite{mukhopadhyay2012} In a previous work, DTN and BPCB were indeed found to show equivalent spin dynamics in the quantum critical regime [seen through the nuclear magnetic resonance (NMR) $T_1$ relaxation time].\cite{mukhopadhyay2012} However, the correctness of the TLL approach for a system which is less 1D than the above-mentioned two spin-ladder compounds has not been specifically addressed so far, and was only very recently discussed regarding NMR relaxation (dynamic) properties.\cite{dupont2016} This question is here addressed by an NMR study of static properties of DTN, in the field close to $H_\textrm{c2}$ (the transition to the fully polarized phase), in which:
(1) we have accurately determined the \textit{absolute} value of the order parameter (that is, the transverse, AF spin component $S_\perp$) in the BEC phase; (2) our low temperature (down to 40~mK) NMR data for the $T$-$H$ phase boundary are compatible with the theoretically expected behavior for the zero-$T$ limit, $T_\textrm{c}(H) \propto (H_\textrm{c2}-H)^{2/3}$, in contrast to previously published results;\cite{yin2008} and (3) we also determined the sample magnetization at $T_\textrm{c}$ and thus the critical boson density, complementing the already existing data close to $H_\textrm{c1}$.\cite{filho2009prl} The complete set of experimental data is accurately reproduced by numerical quantum Monte Carlo (QMC) simulations for the standard 3D $S=1$ model of Eq.~(\ref{Hamiltonian}). This method is computationally quite demanding, but provides a reliable basis to discuss the validity of simpler (approximate) analytical predictions, within either TLL theory or a mean-field Hartree-Fock-Popov (HFP) description.\cite{nikuni2000} Unlike in the previously studied spin ladders, where the 3D (interladder) couplings were taken as a free adjustable parameter to fit the experimental $T_c$ values, in DTN these couplings have been determined independently,\cite{zvyagin2007, tsyrulin2013} making for the theoretical description fully constrained and able to predict the absolute values of observables.

\begin{figure}[t!]
\includegraphics[width=1.00\columnwidth,clip,angle=0]{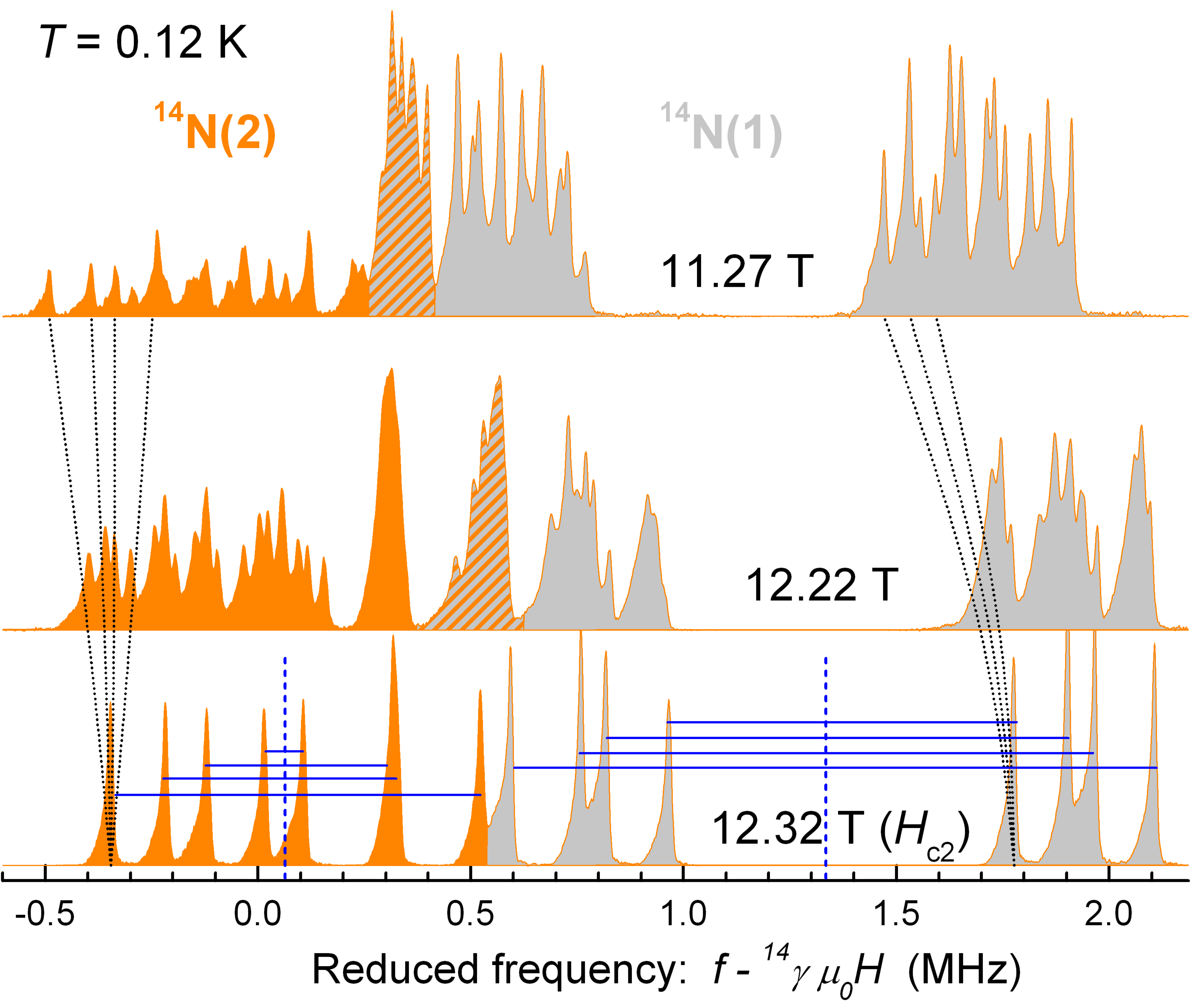}
\caption{Evolution of $^{14}$N NMR spectrum when entering the BEC phase in DTN at $T=0.12$~K. In the spectrum taken at $H_\textrm{c2}=12.32$~T (bottom), the contributions from the two crystallographic sites, orange-colored N(2) and gray-colored N(1), are clearly separated, and the quadrupolar splittings, indicated by blue horizontal lines, are easy to identify. Within the ordered phase, at 12.22 and 11.27~T, the AF order splits each line into 4, as shown on the lowest frequency N(2) line by the dotted curves. Orange-gray hatching denotes the region where the N(2) and N(1) lines overlap. }
\label{spectrafig}
\vspace{-3mm}
\end{figure}

\begin{figure}[t!]
\includegraphics[width=1.00\columnwidth,clip,angle=0]{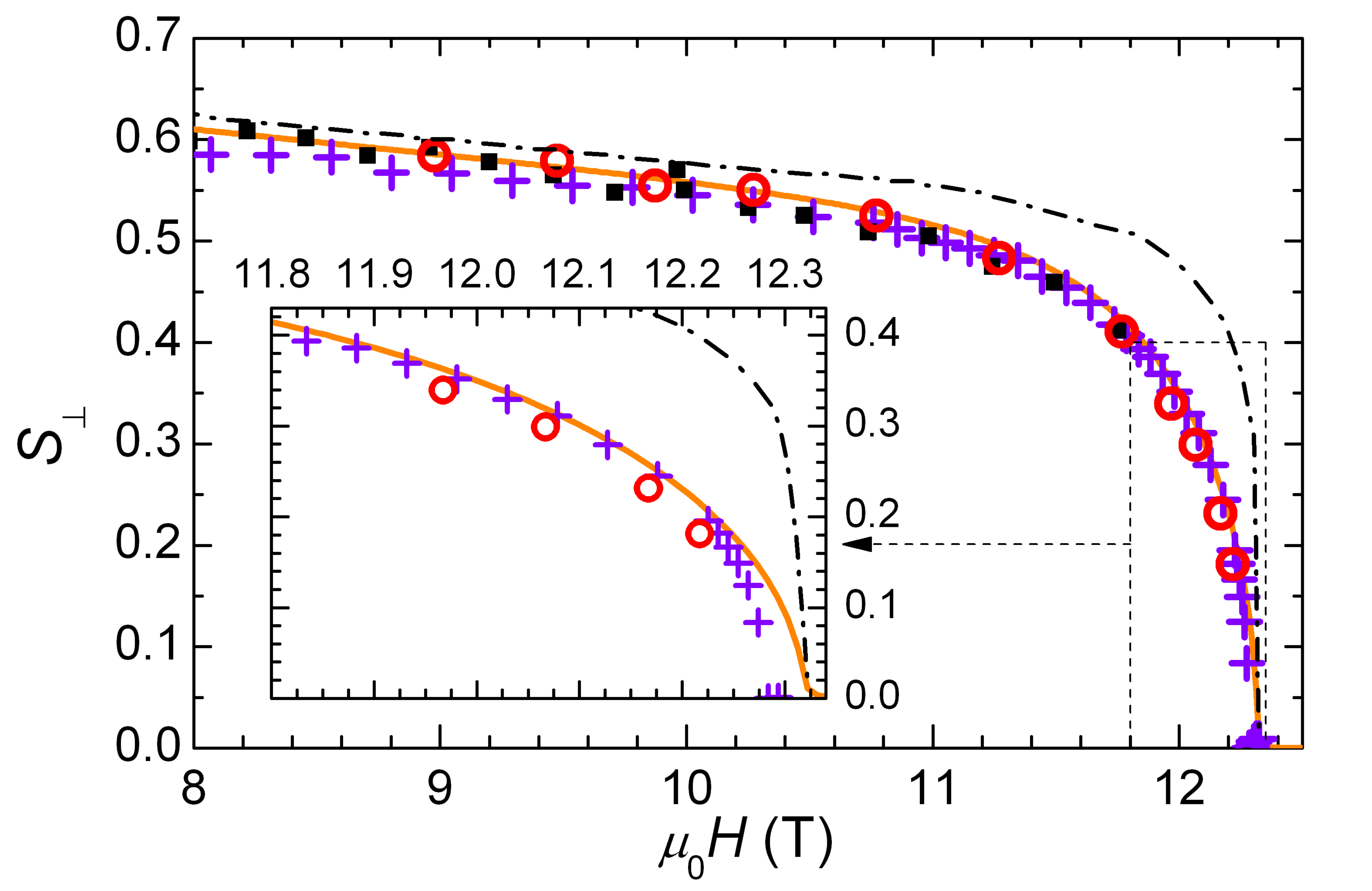}
\caption{The order parameter ($S_\perp$) in the BEC phase of DTN at $T=0.12$ K determined by NMR (circles) and by QMC simulations (crosses). NMR points are overlapped by neutron diffraction data from Ref.~\onlinecite{tsyrulin2013} downscaled by -25\% (squares). The orange solid line and dashed-dotted black curve are the $T=0$ prediction by DMRG+MF and by TLL+MF, respectively. The inset shows a zoom close to $H_\textrm{c2}$. }
\label{ordprmfig}
\vspace{-3mm}
\end{figure}

Experiments were performed on a DTN single crystal of dimensions $\sim 2 \times 2 \times 3$ mm$^3$, placed inside the mixing chamber of a dilution refrigerator, by NMR of proton $^1$H (nuclear spin $I=1/2$) and nitrogen $^{14}$N ($I=1$) nuclei. The local magnetization (spin polarization) of magnetic (Ni$^{2+}$) ions, polarized by the applied magnetic field $H$, is ``seen'' by NMR nuclei as an additional local field $\delta H$ and the corresponding NMR frequency $f=\gamma\mu_0|H + \delta H|/(2\pi)$, where $\gamma$ is the gyromagnetic factor.\cite{slichter3rdedn, abragam} The observed asymmetric lineshape of each individual line in the NMR spectrum (Fig.~\ref{spectrafig}) is well explained by the inhomogeneity of the demagnetizing field over the sample volume.\cite{blinderthesis} $^{14}$N nuclei, in addition, experience the so-called \textit{quadrupolar} coupling to the local electric field gradient (EFG) tensor~\cite{slichter3rdedn, abragam} which strongly splits each NMR line in two (Fig.~\ref{spectrafig}). This splitting has dramatic variations when the sample is rotated, thus allowing precise \textit{in situ} determination of the complete EFG tensor and consequently of the sample orientation (for details, see Ref.~\onlinecite{blinderthesis}). The $c$ axis of the sample was here tilted by $\theta=3.1^{\circ}$ from the field direction.
In order to determine the (nearly) zero-temperature value of the order parameter in the BEC phase, nitrogen spectra were recorded at $T=0.12$~K, a temperature 10 times lower than the maximum $T_\textrm{c}$, $T_\textrm{c}^\textrm{max}=1.2$~K, and at different magnetic fields 9~T $< H < H_\textrm{c2}$ (Fig.~\ref{ordprmfig}). Two different $^{14}$N signals are observed in the NMR spectrum, attributed to the two nitrogen crystallographic sites N(1) and N(2).\cite{figgis1986, blinderthesis,SM} On entering the BEC phase, one can observe that the relative intensity of the N(2) lines is decreasing, which is just an artifact of the measurement sequence (effect of the ``$T_1$'' relaxation).\cite{noteT1} More importantly, a transverse spin component $S_\perp$ appears, corresponding to the BEC order parameter. Since $S_\perp$ is AF ordered, it creates a staggered local field at the nuclei, which results in a splitting of each NMR line. In canonical systems, AF order induces a \textit{doubling} of the unit cell, which results in a splitting of each line into 2, since the local field takes only 2 possible values. This is seen in previous NMR studies.\cite{klanjsek2008, jeong2013} The situation is somewhat more complex in DTN, which has a body-centered tetragonal (BCT) lattice, corresponding to \textit{two} interpenetrating tetragonal subsystems shifted by half of the tetragonal unit cell.\cite{figgis1986} As a result, each NMR line splits into 2$\times$2 = 4 lines when entering an AF ordered phase (see Supplemental Material  (SM)~\cite{SM}). This line splitting is very difficult to follow in the proton spectra (not shown), because they comprise many overlapping lines, but can be successfully tracked in $^{14}$N spectra. Indeed, as soon as $H$ is slightly misaligned from the $c$ axis of tetragonal symmetry, $^{14}$N NMR lines are well separated by the quadrupolar effects (Fig.~\ref{spectrafig}), so that the overlap of lines remains tractable.

To convert the observed line splitting into an order parameter, the main issue is to infer hyperfine tensors \textbf{A} relating the spin polarization $S_{\perp}$ to the observed $\delta H$. For a \textit{homogeneous} order, the corresponding \textbf{A}$(\boldsymbol{q}=0)$ at zero wavevector is easily determined from the NMR lineshift recorded above $H_\mathrm{c2}$, where the system is completely polarized.\cite{blinderthesis} However, the order parameter $S_\perp$ corresponds to $\mathbf{A}(\boldsymbol{q} = \boldsymbol{q}_\textrm{AF}$) at the AF wavevector, whose determination is highly nontrivial (see SM~\cite{SM} for further details). As for the N(1) site further complications are brought by the strong isotropic component of the $\mathbf{A}$ tensor, we decided to quantitatively analyze only the N(2) data, to finally obtain the \textit{absolute value} of the order parameter \textit{amplitude}. Figure~\ref{ordprmfig} shows these results in comparison with different numerical and theoretical descriptions. The magnetic field dependence of $S_\perp$ observed by NMR is perfectly consistent with what was previously reported from neutron measurements (carried out up to 12 T).\cite{tsyrulin2013}  However, in these latter data there is apparently a problem with the determination of the absolute value of $S_{\perp}$, and we had to downscale them by -25\% to make them consistent with the theoretical prediction and the $S_{\perp}$ values determined by NMR.

QMC simulations were performed for a simple tetragonal lattice, considering the two tetragonal subsystems of DTN as totally decoupled. In reality, the BCT crystal structure of DTN contains two interpenetrating tetragonal subsystems, connected to each other through an additional, geometrically \textit{frustrated} coupling ($J_\textrm{f}$)~\cite{tsyrulin2013} that cannot be treated by these simulations. However, corrections to this approximation are expected to be very small, much smaller than the experimentally determined coupling constant, $J_\textrm{f}=80$~mK,\cite{tsyrulin2013} because a perfectly frustrated coupling between the two subsystems  should have no effect at all at the mean-field level. Indeed, despite the approximation, the QMC simulations (see SM~\cite{SM} for more details) performed at $T=0.12$~K agree remarkably well with the experiments.

For a quasi-1D system, a second, computationally less demanding approach to describe the order parameter at $T=0$, is to take into account the interchain (3D) coupling within a mean-field (MF) approximation, neglecting spin fluctuations. This leads to a model of a single DTN chain in an effective magnetic field having a transverse staggered component due to AF transverse ordering and a longitudinal component due to the magnetization. This model can be exactly solved numerically, in a self-consistent way, using the matrix-product state formalism and density matrix renormalization group (DMRG) algorithm,\cite{white1992, white1993, itensor, bouillot2011} to find the ground-state ($T=0$) order parameter. The values obtained in this way, shown in Fig.~\ref{ordprmfig}, are very close to those from QMC (calculated at 0.12~K). Note that the apparent overestimate of $S_\perp$ by this DMRG+MFmethod, of about 3\%, is partly due to the difference in the corresponding temperatures.
Another possibility to solve this 1D effective model is to use analytical expressions based on an (approximate) TLL approach that includes most relevant bosonized MF terms in the Hamiltonian, leading to\cite{giamarchiqphys1D_book, giamarchi1999, klanjsek2008, bouillot2011}
\begin{equation}
  S_\perp = F(K) \sqrt{2A_x} \left( \frac{\pi Z J_{ab} A_x}{u} \right)^{1/(8K-2)},\\
\end{equation}
where $K$ is the TLL exponent, $u$ is the velocity of the excitations, $A_x$ is the amplitude of the transverse correlation function,\cite{noteAx} $Z=4$ is the coordination number along the transverse $a,b$ directions, and $F(K)$ is
\begin{equation}
F\left(K\right) = \left\{\frac{\frac{8K\pi^2}{\left(8K-1\right)\sin\left(\frac{\pi}{8K-1}\right)}\Big[\frac{\Gamma\left(1-\frac{1}{8K}\right)}{\Gamma\left(\frac{1}{8K}\right)}\Big]^{\frac{8K}{8K-1}}}{\left[\Gamma\left(\frac{4K}{8K-1}\right)\Gamma\left(\frac{16K-3}{16K-2}\right)\right]^2}\right\}^{\frac{8K-1}{8K-2}}.\\
\end{equation}
The parameters $K$, $u$ and $A_x$ were obtained as functions of the magnetization $S_z$ using DMRG on a single isolated chain by fitting the transverse correlation function (to get $K$ and $A_x$\cite{hikihara2004}) and by deriving the magnetization curve to get the static susceptibility, $\chi = K/\pi u$, and thereby $u$ (these TLL parameters are shown in SM~\cite{SM}). In Fig.~\ref{ordprmfig} we see that the TLL $S_\perp(T=0)$ values considerably deviate from the DMRG+MF results, especially close to $H_\mathrm{c2}$, while the two results should be identical in the $J_{ab}/J_{c}\ll 1$ limit. Apparently, as regards $S_\perp(T=0)$, the value $J_{ab}/J_{c}\cong 1/12$ is not small enough to consider DTN as a system of \textit{weakly} coupled spin chains and fully justify the TLL description, in particular when the \textit{total} interchain coupling $Z J_{ab}$ becomes larger than the intrachain energy scale $u$
close to $H_\mathrm{c2}$. This is to be compared with other, more 1D compounds, such as BPCB, where the analytical TLL description of $S_\perp$\cite{klanjsek2008, bouillot2011} was remarkably accurate and fully consistent with the numerical (DMRG+MF) treatment.

One may also wonder what is the orientation of the $S_{\perp}$ vector within the $a$-$b$ plane (i.e., the phase of the complex order parameter) on each of the two tetragonal subsystems. When fitting the observed $^{14}$N(2) NMR line splitting in the BEC phase (Fig.~\ref{ordprmfig}) we found several solutions for the orientation, which could not be clearly distinguished.\cite{SM} It turns out that for these different solutions the calculated proton spectra are quite different, and by  comparing them to the experimentally observed one (not shown) we could determine the orientation of the $S_\perp$ vectors.\cite{blinderthesis, SM} On both tetragonal subsystems $S_\perp$ is found to be approximately (within $\sim$10$^\circ$) aligned perpendicularly to the in-plane component of the magnetic field (induced by the 3$^\circ$ tilt of $H$ from the $c$ axis). This orientation is indeed what is expected from the simplest model of classical spins subject to an easy-plane anisotropy. This means that the tilt of the field is the strongest source of axial symmetry breaking. Therefore, for a perfect sample alignment ($H \parallel c$ axis), we do expect that the symmetry requirement for the existence of a true BEC is fulfilled.

\begin{figure}[t!]
\includegraphics[width=1.00\columnwidth,clip,angle=0]{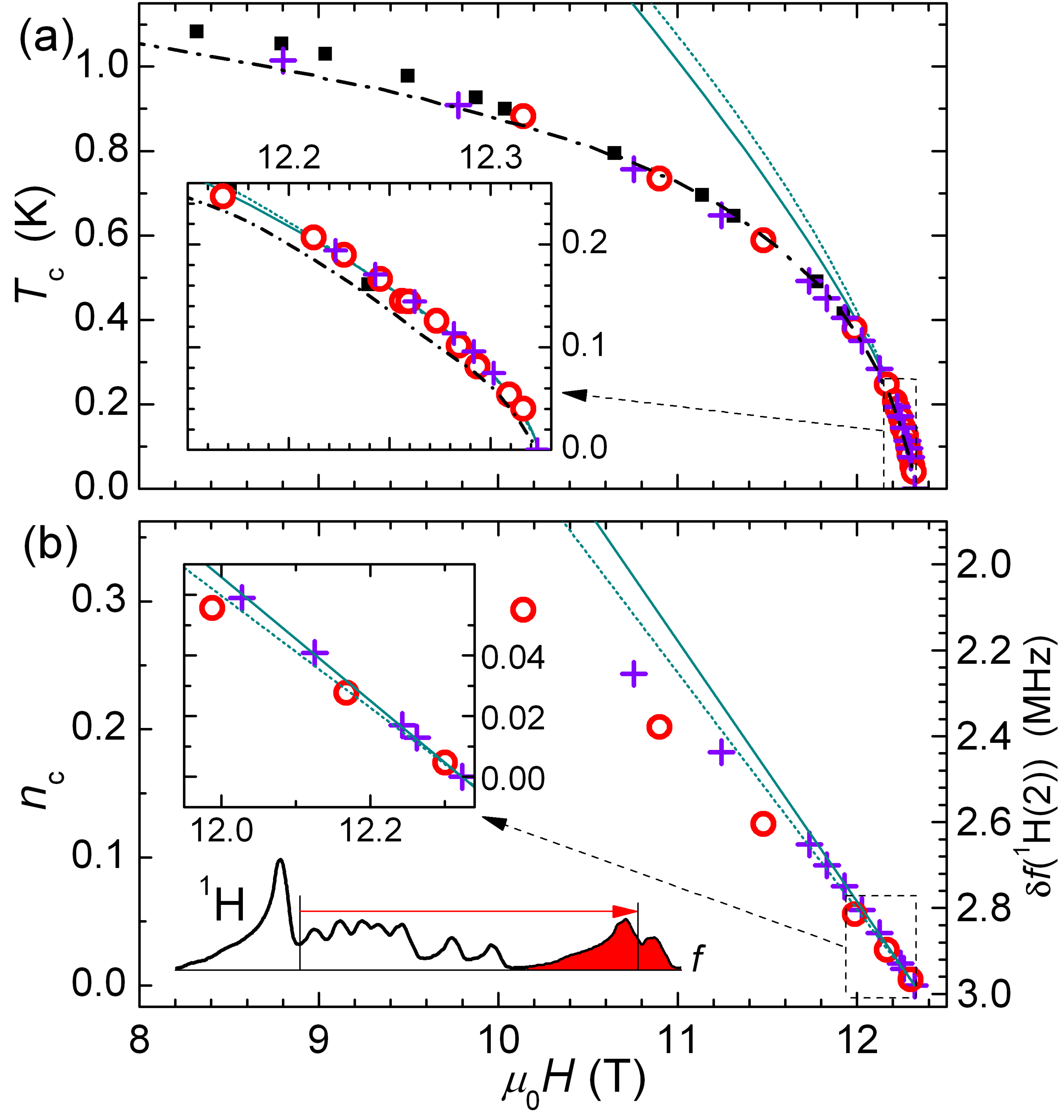}
\caption{NMR data (red circles) for (a) $T_\mathrm{c}$ and (b) the critical boson density $n_\mathrm{c}$ at $T_\mathrm{c}$ (see the text) are compared with theoretical predictions: QMC data points are shown as crosses, HFP  predictions are given by the dotted and solid lines (for the parabolic and true magnon dispersion, respectively), and the TLL prediction for  $T_\mathrm{c}$ [Eq.~(\ref{eqTLLTc}), with $k=0.67$] by the dash-dotted line (see the text). In (a) squares correspond to the magnetocaloric effect data from Ref.~\onlinecite{zapf2006} (with the field values downscaled by -2.3\% to overlap the slightly different  $H_\textrm{c2}$ values). The lower inset in (b) explains the determination of the H(2) line-shift frequency (right scale in the main panel) that measures $n_\mathrm{c}$. Other insets are zooms close to $H_\textrm{c2}$.}
\label{densityTcfig}
\vspace{-5mm}
\end{figure}

We now turn our attention to the \textit{phase boundary} $T_\mathrm{c}(H)$ (Fig.~\ref{densityTcfig}). By NMR the precise $T_\mathrm{c}$ value is detected from the peak position of the corresponding critical spin fluctuations. We measured it on the high-frequency proton $^1$H(2) line as the corresponding peak of the transverse nuclear spin-spin relaxation rate (1/$T_2$) or, equivalently, as the minimum intensity of the NMR signal recorded while varying $H$ or $T$ through the transition. From the same set of proton NMR measurements we also extracted the longitudinal spin component $S_z$ \textit{at the transition}, $S_z(T_\mathrm{c})$. This quantity, which relates to the critical boson density $n_\mathrm{c}$ (assuming holelike bosonic particles), $n_\mathrm{c} = 1 - S_z(T_\mathrm{c})$, was accessed by the frequency shift of the $^1$H(2) line $\delta f =  \gamma  \mathrm{A}_{cc} g_c \mu_\mathrm{B} S_z / (2 \pi) = \widetilde{\mathrm{A}}_{cc} S_z$, where $\mathrm{A}_{cc}$ is the relevant component of the hyperfine tensor (see SM~\cite{SM}, $\widetilde{\mathrm{A}}_{cc} = 2.98$ MHz).

In Fig.~\ref{densityTcfig} we can see that QMC simulations are in excellent agreement with NMR data for both $n_\mathrm{c}$ and $T_\mathrm{c}$ (as was the case for $S_\perp$ in Fig.~\ref{ordprmfig}). $T_\mathrm{c}$ can also be described by the analytical TLL-based expression\cite{klanjsek2008, bouillot2011}
\begin{equation}
T_\mathrm{c} = \frac{u}{2\pi}\left[ \sin\left(\frac{\pi}{4K}\right)B^2 \left( \frac{1}{8K}, 1 - \frac{1}{4K} \right) \frac{k Z J_{ab} A_x}{u}\right]^{\frac{2K}{4K-1}}, \label{eqTLLTc}
\end{equation}
where $B(X,Y) = \Gamma(X)\Gamma(Y)/\Gamma(X+Y)$, except very close to $H_\textrm{c2}$ where the TLL description fails.\cite{giamarchi1999} Here we have explicitly included a renormalization parameter $k$ to take into account the effects of spin fluctuations beyond the MF treatment of interchain interaction. This was first discussed analytically for the Heisenberg  spin chain in zero field in Ref.~\onlinecite{irkhin2000} and then precisely verified numerically in Ref.~\onlinecite{yasuda2005}, where $k=0.695$ was obtained. A slightly different value, $k=0.74$ was successfully applied in describing $T_\mathrm{c}(H)$ of the BPCB compound,\cite{thielemann2009, bouillot2011} while for our DTN data we find $k=0.67$(2), pointing to a quite universal value of this correction.

Close to $H_\textrm{c2}$ one expects that the 3D description of the HFP model, describing the low boson density limit, is valid. Using the low-energy quadratic approximation for the magnon dispersion,\cite{nikuni2000} this model provides the canonical shape of the phase boundary, $T_\mathrm{c}(H) \propto (H_\mathrm{c2} - H)^{2/3}$, which is well observed by our NMR data, in contrast to previous reports.\cite{yin2008} (From a nonlinear power-law fit the exponent value is 0.72\,$\pm$\,0.04.)
To better access higher temperature, one can improve the model by taking the exact, numerically calculated dispersion of magnons (as in Ref.~\onlinecite{misguich2004}), which indeed fits the data slightly better. In both cases, the interaction parameter $U_\mathrm{3D} = g_c \mu_\mathrm{B}(H_\mathrm{c2} - H)/(2 k_\mathrm{B} n_\mathrm{c})$ was fitted to adapt the $T_\mathrm{c}(H)$ data points below 0.25~K.  The obtained values, $U_\mathrm{3D}=4.1$ and 3.7~K, are perfectly consistent with the initial slope of the measured $n_\mathrm{c}(H)$ dependence shown in Fig.~\ref{densityTcfig}(b), confirming the validity of the HFP model. We remark that close to $H_\mathrm{c1}$ a higher value, $U_\mathrm{3D}=7.2$~K, was reported,\cite{filho2009prl} which should be attributed to the renormalization described in Ref.~\onlinecite{Kohama2011}.
In Fig.~\ref{densityTcfig}(a) we also see that the HFP model in both variants clearly fails above 0.3~K, corresponding to $n_\mathrm{c}\cong 4$\%. We have verified that this cannot be compensated by taking the renormalized field-dependent $U_\mathrm{3D}$ value from the observed $n_\mathrm{c}(H)$ dependence, meaning that above $n_\mathrm{c}\cong 4$\% the interactions modify the effective density of states as compared to its noninteracting value.

To conclude, by NMR we investigated static properties of the high-field part of the BEC phase in the \mbox{quasi-1D} quantum magnet DTN, and analyzed the data using several theoretical approaches. QMC numerical simulations for a standard spin-1 model provide excellent fit to the data, and we used them as a reference to discuss the applicability of other approximate techniques and their sensitivity to the strength of 3D coupling. For a moderately 1D system such as DTN ($J_{c}/J_{ab}\cong$ 12) we find that analytical TLL-based predictions are still very good for $T_\mathrm{c}$ (when the renormalization of MF interaction is taken into account) but insufficient for the order parameter $S_{\perp}$. For $S_{\perp}$, DMRG+MF turns out to be precise, and does not require any renormalization. The HFP description is found to be valid only very close to $H_\mathrm{c2}$, for the critical boson densities below 4\%.

\begin{acknowledgments}
This work was performed using HPC resources from GENCI (Grant No. x2016050225), and is supported by the French ANR project BOLODISS (Grant No. ANR-14-CE32-0018) and by R\'{e}gion Midi-Pyr\'{e}n\'{e}es.  M.S.G. acknowledges the support of Croatian Science Foundation (HRZZ) under the project 2729, and the Unity through Knowledge Fund (UKF Grant No. 20/15). A.P.-F. acknowledges the support of the Brazilian agencies CNPq and FAPESP (Grant No. \mbox{2015-16191-5}).
\end{acknowledgments}

\clearpage

\renewcommand{\thefigure}{S\arabic{figure}}
\setcounter{figure}{0}
\renewcommand{\theequation}{S\arabic{equation}}
\setcounter{equation}{0}

\begin{center}

\textbf{\large SUPPLEMENTAL MATERIAL \\} \vspace{15pt}

\textbf{to: ``Nuclear Magnetic Resonance study of the magnetic-field-induced ordered phase in the NiCl$_{2}$-4SC(NH$_{2}$)$_{2}$ compound'' by  R.~Blinder,$^*$  M.~Dupont, S.~Mukhopadhyay,$^\dag$ M.~S.~Grbi\'{c}, N.~Laflorencie, S.~Capponi, H.~Mayaffre, C.~Berthier, A.~Paduan-Filho, and M. Horvati\'c$^\ddag$}

\end{center}

\section{Quantum Monte Carlo simulations}

Throughout the paper we use quantum Monte Carlo (QMC) stochastic series expansion (SSE) algorithm.\cite{Ssandvik1999, SALPS}
We work with finite size systems counting \mbox{$N = L \times L/8 \times L/8$} spins where $L$ is the number of consecutive $S=1$ spins along the chains ($c$) direction. In order to adapt the simulation to the 1D character of the Hamiltonian, only $L/8$ spins are then taken in each transverse ($a$ and $b$) directions, meaning an aspect ratio of $1/8$.



\subsection{Transverse order parameter}

\begin{figure}[ht!]
    \includegraphics[width=.865\columnwidth,clip,angle=0]{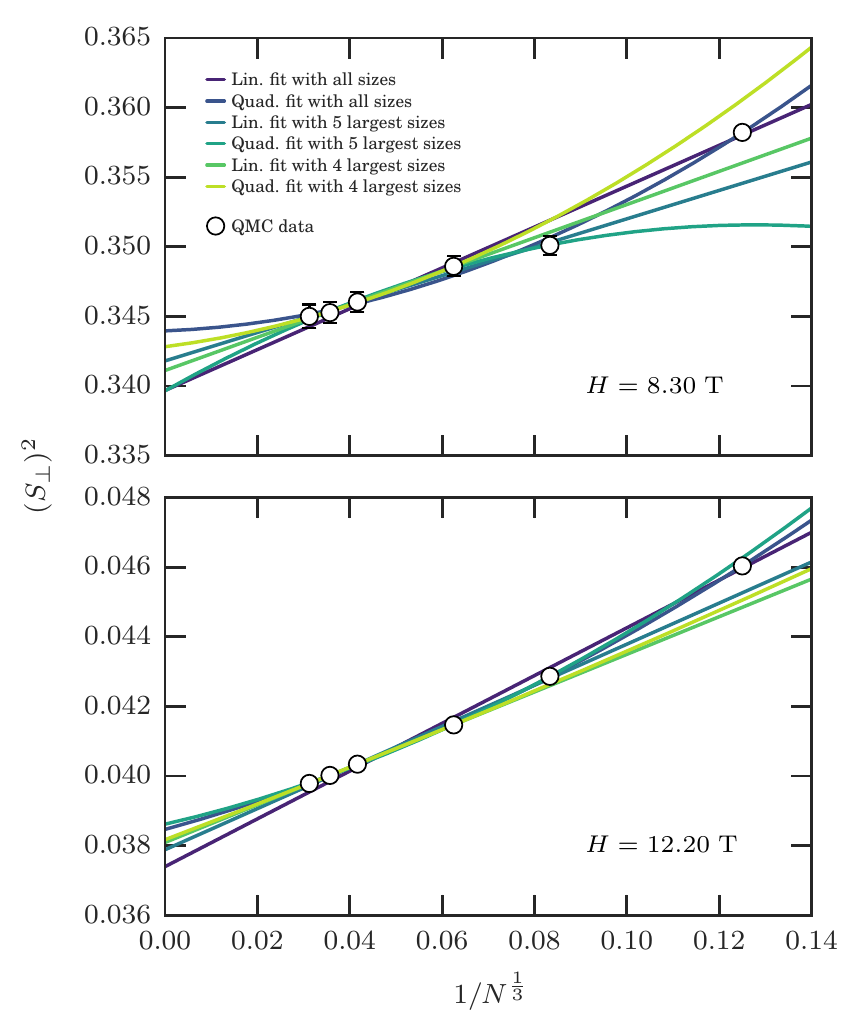}
    \caption{Two examples, at different magnetic fields, for the extrapolation of the order parameter to the thermodynamic limit. Various linear and quadratic fits (solid lines) are performed taking into account a varying number of QMC data points (circles). SSE simulations are performed for the DTN model at $T=0.12$~K.}
     \label{sperp_scaling}

\end{figure}

To reliably extract the \textit{order parameter} $S_{\perp}$ at the thermodynamic limit, we performed simulations for different system sizes, up to $L=128$, and did various linear and quadratic fits of the transverse structure factor
\begin{equation}
    \left(S_{\perp}\right)^2=\frac{1}{2N^2}\sum_{i,j}e^{i\boldsymbol{q}\cdot(\boldsymbol{r}_i-\boldsymbol{r}_j)}\langle S_i^+ S_j^- + S_i^- S_j^+\rangle
    \label{eq:mperp_square}
\end{equation}
at $\boldsymbol{q}=(\pi,\pi,\pi)$ as a function of $1/N^\frac{1}{3}$ (Fig.~\ref{sperp_scaling}), which allows to extrapolate the value to the thermodynamic limit ($N\rightarrow\infty$). The summation in Eq.~\eqref{eq:mperp_square} is over all possible sites $i$ and $j$ of the lattice. The values of the order parameter given in Fig.~2 of the main manuscript correspond to the square root of the mean value coming from the various extrapolations of $(S_{\perp})^2$, while the error bars correspond to the standard deviation of these extrapolations around their mean value. Thus they do not directly reflect the QMC errors even though they have been taken into account when performing the fits.

\subsection{Critical temperature}

\begin{figure}[h!]
    \includegraphics[width=.89\columnwidth,clip,angle=0]{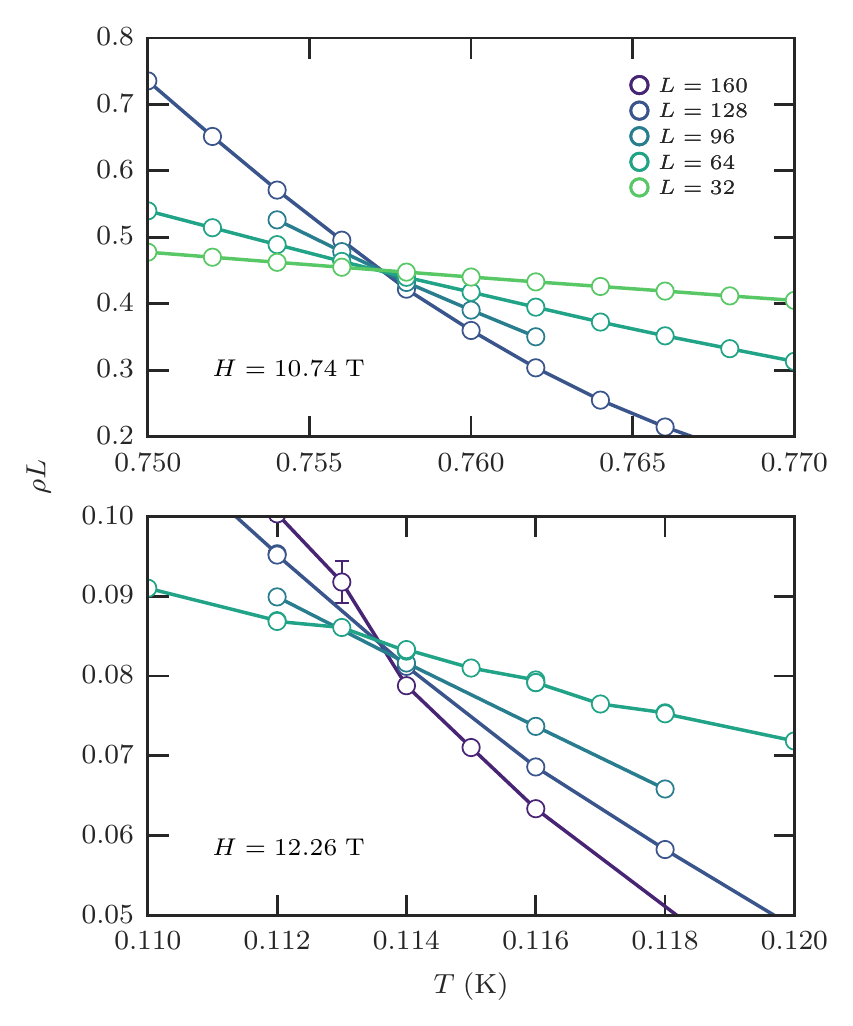}
    \caption{Two examples, at different magnetic fields, of the crossing of the spin stiffness $\rho$ times the system size $L$ to determine the critical temperature.}
     \label{stiffness_crossing}
\end{figure}

The critical temperature was numerically determined using QMC through the crossing of the spin stiffness $\rho$ times $L$ for various sizes up to $L=160$. Indeed, at the critical point one expects a scaling ansatz for $\rho$ which only depends on the dimensionality ($d=3$ in the case of DTN) such that $\rho L\sim$~constant (Fig.~\ref{stiffness_crossing}). More precisely, the estimated values of $T_\mathrm{c}(H)$ given in Fig.~3, as well as their error bars, are determined performing Bayesian scaling analysis \cite{Sharada1, Sharada2} of the spin stiffness data.


\section{Tomonaga-Luttinger liquid (TLL) parameters}

\begin{figure}[h!]
\includegraphics[width=1\columnwidth,clip,angle=0]{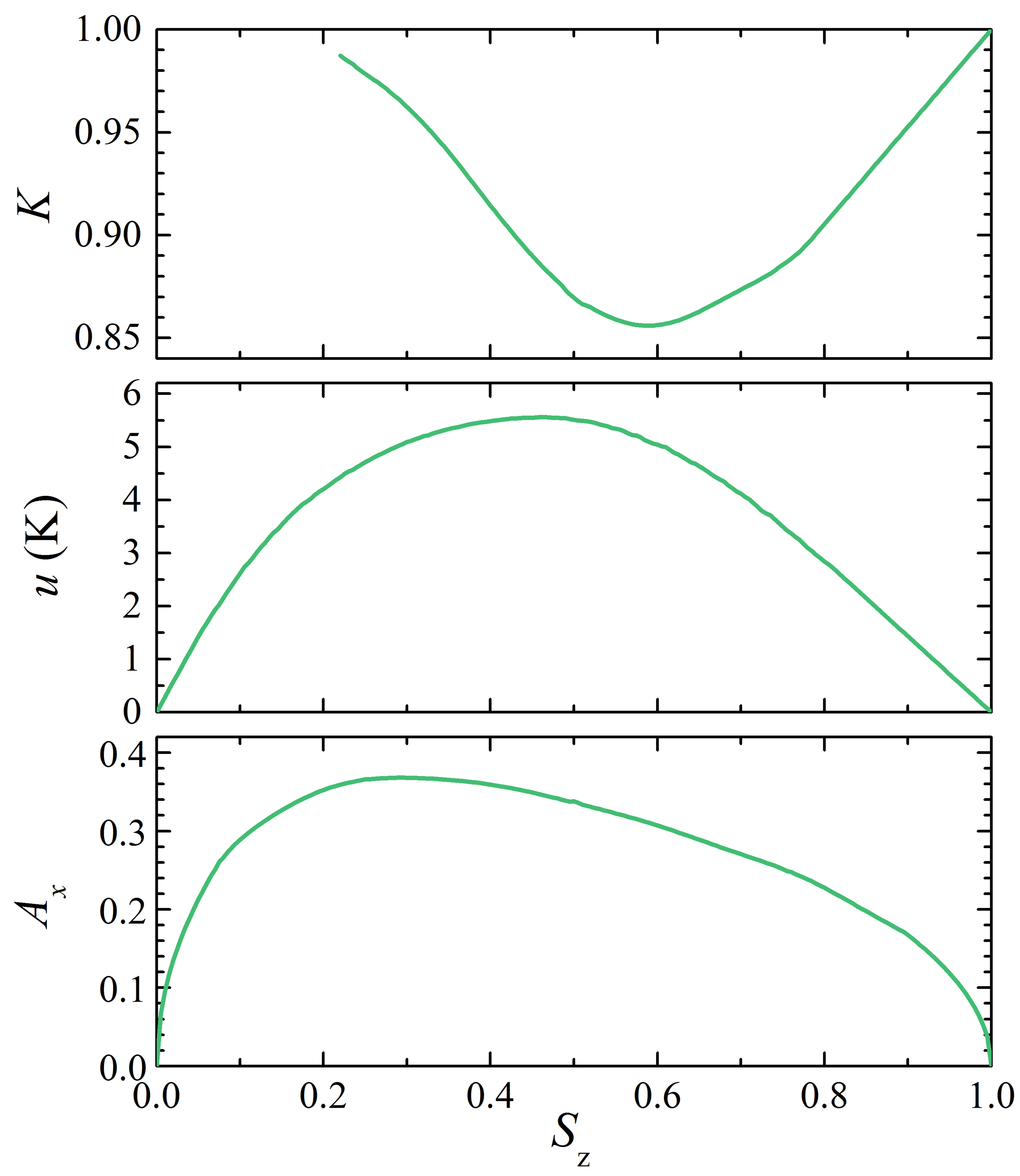}
\caption{TLL parameters as a function of sample magnetization $S_z$, calculated by DMRG for the DTN ($J_c=2.2$~K, $D=8.9$~K). Values for $K$ at $S_z \rightarrow 0$ are not represented as the behavior of $K$ in that region could not be precisely defined. }
 \label{LLparamsfig}
\vspace{-3mm}
\end{figure}

Fig.~\ref{LLparamsfig} shows the TLL parameters for the DTN, determined numerically by DMRG. This determination is increasingly difficult when approaching critical fields, in the $S_z \rightarrow 0$ and $S_z \rightarrow 1$ limits.
These values are thus better defined from the expected analytical behaviors: $u(S_z)$ and $A_x(S_z)$ are approaching zero linearly and as a square root, respectively. $K(S_z)$ is linearly approaching 1 on both sides, but its $S_z \rightarrow 0$ behavior could not be precisely defined.

\begin{figure}[h!]
\includegraphics[width=0.97\columnwidth,clip,angle=0]{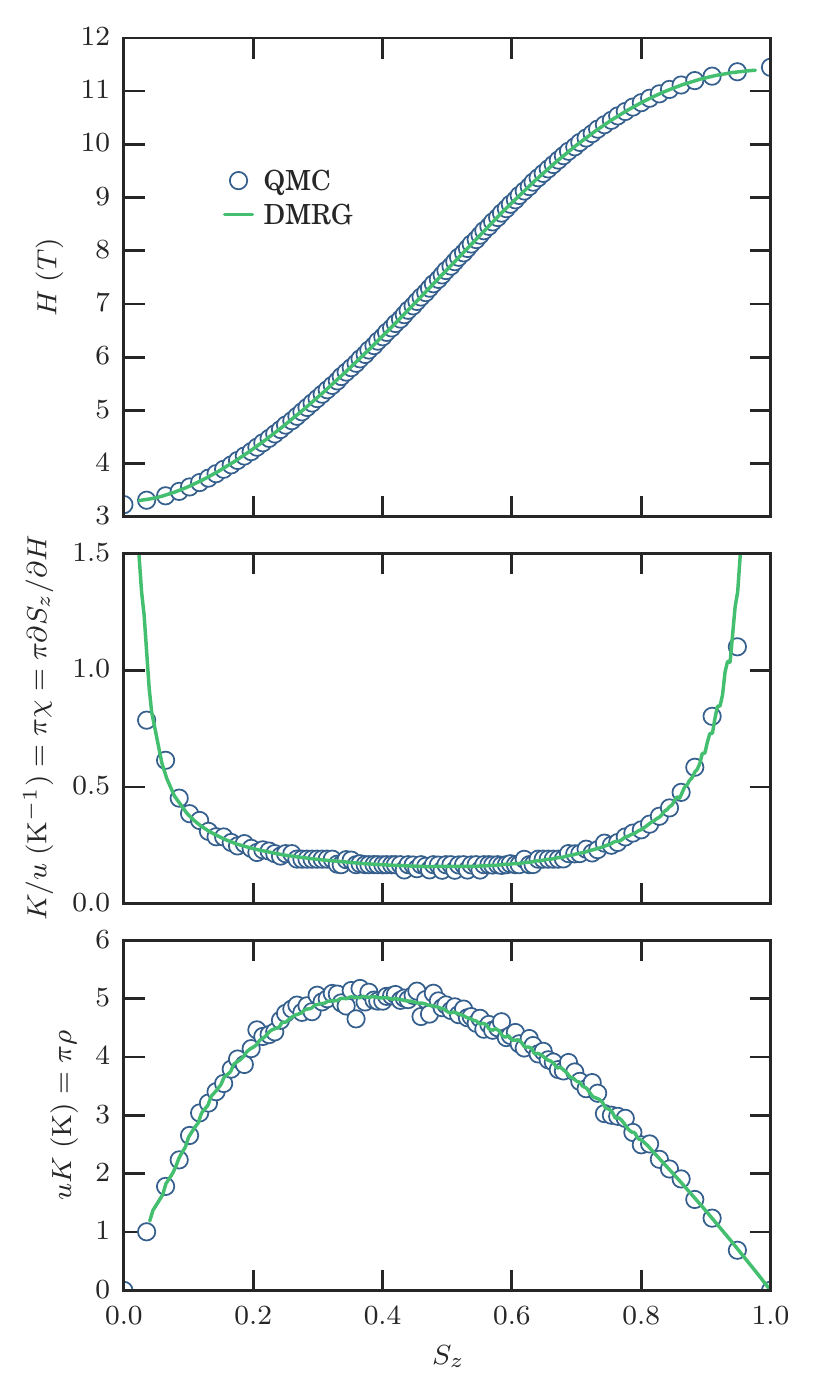}
\caption{Comparison between QMC (circles) and DMRG (lines) results to determine the values of $u$ and $K$. The susceptibily $\chi$ is the first derivative of the magnetization curve $S_z(H)$ and $\rho$ is the spin stiffness. QMC error bars are smaller than the symbol size and thus not visible. Note that the critical fields observed in the first panel are purely one dimensional.}
 \label{tll_parameters}
\vspace{-3mm}
\end{figure}

In order to validate the values computed with DMRG (or in an attempt to provide better estimates), independent QMC simulations using SSE algorithm were carried out on a single DTN chain ($J_c=2.2$~K, $D=8.9$~K) containing $L=512$ spins. Simulations were performed at fixed magnetic field and low enough temperature (\mbox{$T=1/2L$} kelvins) to ensure we are probing the ground state. Fig.~\ref{tll_parameters} shows the relevant thermodynamic quantities: the longitudinal magnetization $S_z$ and the spin stiffness, which can be related to the TLL parameters $u$ and $K$. The QMC results are indeed identical to DMRG values, but they do not provide more precise estimate of the TLL parameters.

\section{Hyperfine coupling tensor}

\begin{table}[t!]
\setlength\tabcolsep{8pt}  
\begin{tabular}{|c|c|c|c|}
\hline                                                              & $^1$H(3)      & $^1$H(4)  & $^1$N(2)  \\  \hline
$\mathrm{A}_{ca}(\boldsymbol{q} = \boldsymbol{0})$    NMR spectra &\bf{23.7}  & \bf{33.5} & \bf{22.3}     \\
\hspace{36pt}                                       Simulation  &  24.0     &   37.7    & 21.2      \\ \hline
$\mathrm{A}_{cb}(\boldsymbol{q} = \boldsymbol{0})$    NMR spectra & \bf{23.3} & \bf{32.3} & \bf{20.6}     \\
\hspace{36pt}                                       Simulation  &  23.5     &   36.4    & 24.4   \\ \hline
\end{tabular}
\caption{The $\boldsymbol{q}=\boldsymbol{0}$ off-diagonal components of the hyperfine coupling tensor for selected crystallographic sites in DTN, all in mT/$\mu_\mathrm{B}$ units. We present both the values determined from the rotational dependence of the NMR spectra and the predicted dipolar values, calculated using crystallographic positions from Ref.~\onlinecite{Sfiggis1986}.}
\label{tabAoffdiag}
\end{table}

\begin{table}[h!]
\vspace{0.4cm}
\setlength\tabcolsep{6pt}  
\begin{tabular}{|c|c|c|c|c|c|c|}
\hline
                & $^1$H(1)  & $^1$H(2)  & $^1$H(3)  & $^1$H(4) & $^{14}$N(1) & $^{14}$N(2) \\ \hline
$\mathrm{A_{iso}}$    & -0.3      &   20.9    & 3.9       & 4.1      & 207        & 15.8 \\ \hline
\end{tabular}
\caption{Isotropic component of the hyperfine coupling tensor for all proton and nitrogen sites in DTN, in mT/$\mu_\mathrm{B}$ units. These components were measured from rotational dependence of the spectra in the fully polarized regime.}
\label{tabAiso}
\end{table}

By NMR we measure the local hyperfine field $\delta H$ induced at a given nuclear position by the surrounding magnetic moments. Linear relation connecting measured $\delta H$ to the corresponding local spin values $\boldsymbol{S}(\boldsymbol{r})$ at different positions $\boldsymbol{r}$,
\begin{equation}
\mu_0 \delta \boldsymbol{H} = \sum\limits_{\boldsymbol{r}} \mathbf{A}(\boldsymbol{r}) g \mu_\mathrm{B} \boldsymbol{S}(\boldsymbol{r}),
\end{equation}
defines the (local) hyperfine coupling tensor\cite{Sabragam,Sslichter3rdedn}  $\mathbf{A}(\boldsymbol{r})$ (where the $g$~tensor is diagonal in the $a,b,c$ crystallographic frame of DTN and amounts to $g_c=2.26$ and $g_{ab}=2.34$\cite{Sfilho1981,Szvyagin2007}).
Experimental determination of the $\mathbf{A}$ tensor is a ``calibration'' that is necessary to convert measured NMR line-shift frequency into the absolute value of spin-polarization.
The $\mathbf{A}(\boldsymbol{r})$ tensor can be written as a sum of two contributions:
\begin{equation}
\mathbf{A}(\boldsymbol{r}) = \mathbf{A}(\boldsymbol{r})^{\mathrm{dip}} + \delta \mathbf{A}(\boldsymbol{r}),
\end{equation}
where $\mathbf{A}(\boldsymbol{r})^{\mathrm{dip}}$ is the known dipolar term (decreasing with distance as $1/r^3$), which we calculate using the crystallographic positions from Ref.~\onlinecite{Sfiggis1986} and assuming the spins to be fully localized (point-like) at the magnetic sites. $\delta \mathbf{A}$ is a priori unknown deviation from the purely dipolar contribution, which can be attributed to neighboring spins. It can originate from different effects, such as delocalization of the spin towards the sites neighboring the nickel ions, from a transferred hyperfine term if the valence orbitals surrounding the considered nucleus are themselves subject to polarization transfer, \cite{Sabragam, Sslichter3rdedn} or, in the case of proton, as an artefact of incorrectly known crystallographic positions.  In order to measure the $\delta \mathbf{A}$ component, experiments were performed using a variable temperature insert at $H=15$~T, $T=1.5$~K, \textit{in the fully polarized regime} where $|\boldsymbol{S}(\boldsymbol{r})|=1$. In this regime, we measured the $^1$H and $^{14}$N NMR spectrum as a function of sample orientation. For the proton $^1$H, the sample was tilted up to $\pm15^{\circ}$ from an almost perfectly aligned orientation ($\theta\sim 1^{\circ}$) of magnetic field with respect to $c$ axis. For the nitrogen $^{14}$N, the sample was tilted $-10^{\circ}/+90^{\circ}$ from a quasi-aligned orientation ($\theta\sim 2^{\circ}$). Since in this regime the magnetic moment is homogeneous, only the space-integrated component $\sum_{\boldsymbol{\boldsymbol{r}}} \mathbf{A}(\boldsymbol{r}) = \mathbf{A}(\boldsymbol{q} = \boldsymbol{0})$ of the hyperfine tensor is measured. Electric Field Gradient (EFG) tensor and sample orientation were obtained by fitting the quadrupolar contribution to the lineshift.\cite{Sabragam, Sslichter3rdedn, Sblinderthesis} Regarding the hyperfine coupling, we focus on the off-diagonal terms $\mathrm{A}_{ca}(\boldsymbol{q}=\boldsymbol{0})$ and $\mathrm{A}_{cb}(\boldsymbol{q}=\boldsymbol{0})$ that are important for the analysis of spectra in the BEC phase.
\begin{figure}[t!]
\includegraphics[width=1.00\columnwidth,clip,angle=0]{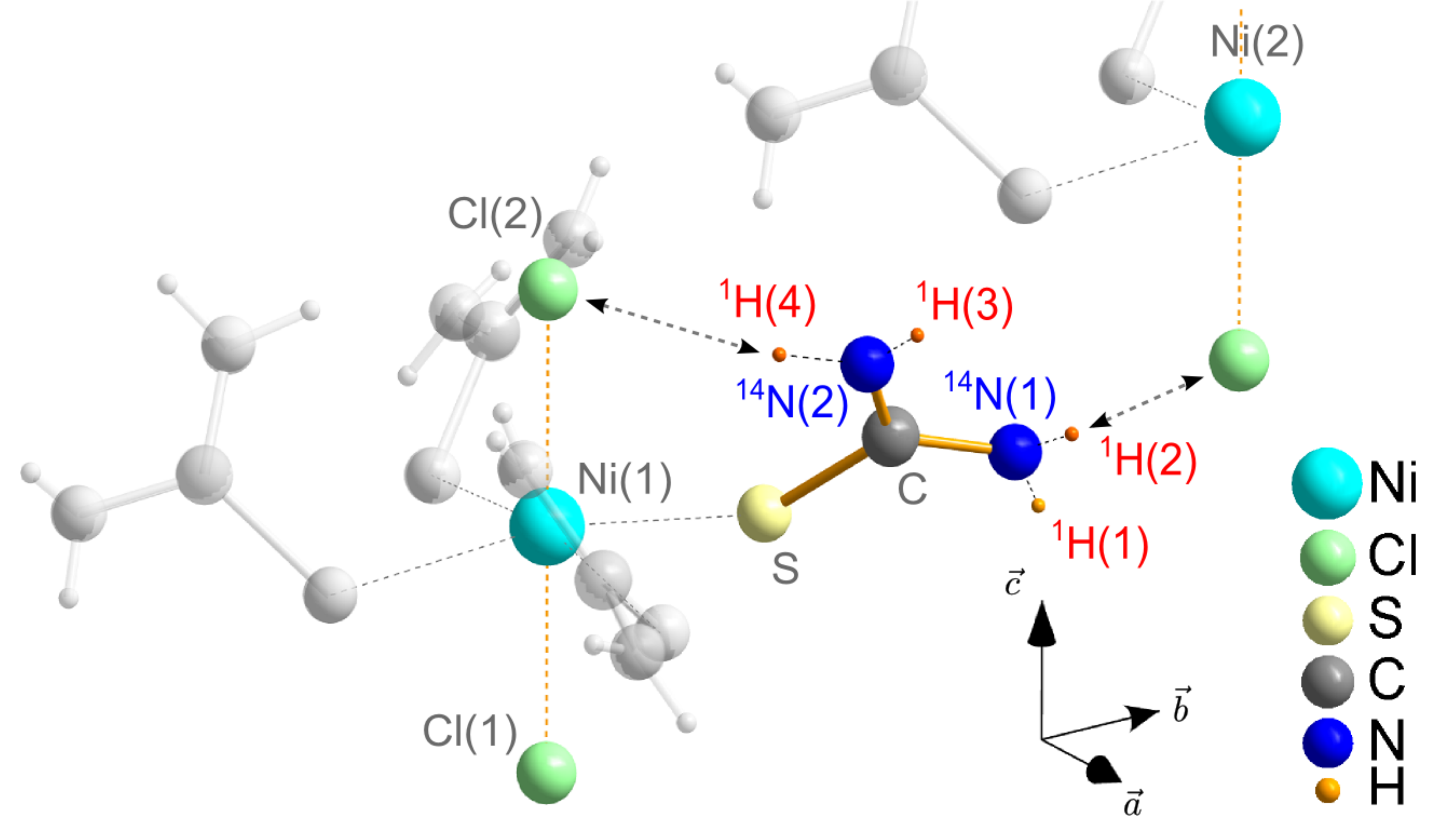}
\caption{Location of the $^1$H (red) and $^{14}$N (blue) NMR sites within the unit cell in DTN. The Ni(1) and Ni(2) nickels seen on this figure belong to the 2 different tetragonal subsystems. Dashed double arrows indicate hydrogen bonds.}
 \label{strucDTNfig}
\vspace{-3mm}
\end{figure}
These terms could be extracted with very good accuracy for sites $^1$H(3), $^1$H(4) and $^{14}$N(2) (defined in Fig.~\ref{strucDTNfig}), see Table \ref{tabAoffdiag}.\cite{Sblinderthesis} Table \ref{tabAoffdiag} presents also the calculated dipolar values for the corresponding sites, showing that measured terms are nearly completely of dipolar origin. For this contribution it is straightforward to calculate the values corresponding to an AF order of spin polarization, $\mathbf{A}(\boldsymbol{q} = \boldsymbol{q}_\mathrm{AF})$, that are relevant for the determination of the order parameter, see the following section.

In Table \ref{tabAiso} we also show the \textit{isotropic} part of the hyperfine coupling, defined as $\mathrm{A_{iso}} = \rm{tr}(\mathbf{A})/3$, for \textit{all} proton and nitrogen sites. They reflect the spin polarization of the orbitals that have non-zero value at the location of given nuclear site, thus giving local information from close neighbors. On the contrary, for an orbital which do not spatially overlap with the nucleus, the through-space dipolar coupling tensor would by definition give $\mathrm{A_{iso}}=0$. We remark that among protons/nitrogens, the most important value of $\mathrm{A_{iso}}$ is found for $^1$H(2)/$^{14}$N(1), which suggests a common polarization transfer path. The bigger $\mathrm{A_{iso}}$ is, the stronger is the coupling to the longitudinal (homogeneous) part of the magnetic moment; relatively high values of $\mathrm{A_{iso}}$ for sites $^1$H(2) and $^{14}$N(1) thus explain why the corresponding lines are shifted to high frequency in the NMR spectra taken within the fully polarized phase (see Fig.~1 and inset of Fig.~3(a)).    

\vspace{-2pt}
\section{Order parameter in the BEC phase}
\vspace{-2pt}

The ordered (BEC) phase is characterized by the AF spin component $S_{\perp}$, corresponding to the BEC order parameter. In canonical systems, the measured staggered local field would be written as \mbox{$\mu_0 \delta \boldsymbol{H}^\textrm{AF} = \mathbf{A}(\boldsymbol{q}_\textrm{AF}) g \mu_\mathrm{B} \boldsymbol{S}_\perp$}, where $\boldsymbol{q}_\mathrm{AF} = (\pi, \pi, \pi)$ is the AF wavevector. However, DTN has the Body-Centered Tetragonal (BCT) lattice, which corresponds to two interpenetrating tetragonal subsystems shifted by half of the tetragonal unit cell. Then, a general expression for the local field reads:
\begin{equation}
\mu_0 \delta \boldsymbol{H}^\textrm{AF} = \epsilon_1 \mathbf{A}^1(\boldsymbol{q}_\textrm{AF}) g \mu_\mathrm{B} \boldsymbol{S}_{\perp 1} + \epsilon_2 \mathbf{A}^2(\boldsymbol{q}_\textrm{AF}) g \mu_\mathrm{B} \boldsymbol{S}_{\perp 2}  ,
\label{eqlocfieldAF}
\end{equation}
where we separate the contributions from the two tetragonal subsystems (denoted by indices 1 and 2), $\epsilon_{1,2} = \pm 1$, and $\boldsymbol{S}_{\perp 1}$, $\boldsymbol{S}_{\perp 2}$ are the corresponding order parameters (where symmetry constrains only their amplitudes, but not directions, to be equal, $|\boldsymbol{S}_{\perp 1}| = |\boldsymbol{S}_{\perp 2}|$). Note that eq. (\ref{eqlocfieldAF}) gives \textit{4 different values of the local field, even if $\boldsymbol{S}_{\perp 1}$ and  $\boldsymbol{S}_{\perp 2}$ are identical}. Therefore, each NMR line is expected to split into 4 lines when entering the BEC phase of DTN.\cite{Sblinderthesis}
    In order to determine precisely the order parameter from the $^{14}$N line-splittings, we need to know the hyperfine couplings to the AF magnetization $\mathbf{A}^1(\boldsymbol{q}_\textrm{AF})$ and $\mathbf{A}^2(\boldsymbol{q}_\textrm{AF})$ in equation \ref{eqlocfieldAF} (more precisely, the off-diagonal $\mathrm{A}_{ca}^1$, $\mathrm{A}_{cb}^1$, $\mathrm{A}_{ca}^2$, $\mathrm{A}_{cb}^2$, components). As already mentioned, these can be easily calculated for the dipolar contribution, while the remaining contribution/correction, $\delta \mathrm{A}_{ca}^{1,2}(\boldsymbol{q}_\textrm{AF})$, should be treated separately and involves additional modeling/assumptions. Regardless of the exact mechanism leading to this contribution, it must be of local type reflecting the state of the neighboring magnetic ions only. As $^{14}$N(2) is located roughly in between two nickels belonging to each tetragonal subsystem (Ni(1) and Ni(2) on Fig.~\ref{strucDTNfig}), we have chosen to assume that the deviation is equally shared between the two subsystems, leading to:
\begin{equation}
\delta \mathrm{A}_{ca,cb}^1(\boldsymbol{q}_\textrm{AF}) = \delta \mathrm{A}_{ca,cb}^2(\boldsymbol{q}_\textrm{AF}) = \delta \mathrm{A}_{ca,cb}(\boldsymbol{q}=\boldsymbol{0})/2
\label{eqAFhyp}
\end{equation}

We remark that our determination of $\boldsymbol{S}_\perp(H)$ is only weakly dependent on the assumptions needed to define the $\delta\mathbf{A}^{1,2}(\boldsymbol{q}_\textrm{AF})$ terms, simply because the hyperfine coupling tensor is strongly dominated by the known dipolar component. For instance, alternatively to Eq.~(\ref{eqAFhyp}), the assumption that only Ni(1) is involved in the local correction to the known dipolar field ($\delta \mathbf{A}^{1}(\boldsymbol{q}_\textrm{AF}) = \delta \mathbf{A}(\boldsymbol{q}=\boldsymbol{0})$, $\delta \mathbf{A}^{2}(\boldsymbol{q}_\textrm{AF}) = 0$) leads to modification of the $S_{\perp}$ value by only +3\%. The ``symmetric'' hypothesis for Ni(2) ($\delta \mathbf{A}^{2}(\boldsymbol{q}_\textrm{AF}) = \delta \mathbf{A}(\boldsymbol{q}=\boldsymbol{0})$, $\delta \mathbf{A}^{1}(\boldsymbol{q}_\textrm{AF}) = 0)$ leads to a symmetric correction of -3\%.

\begin{figure}[h!]
\vspace{0.2cm}
\includegraphics[width=125pt,clip,angle=0]{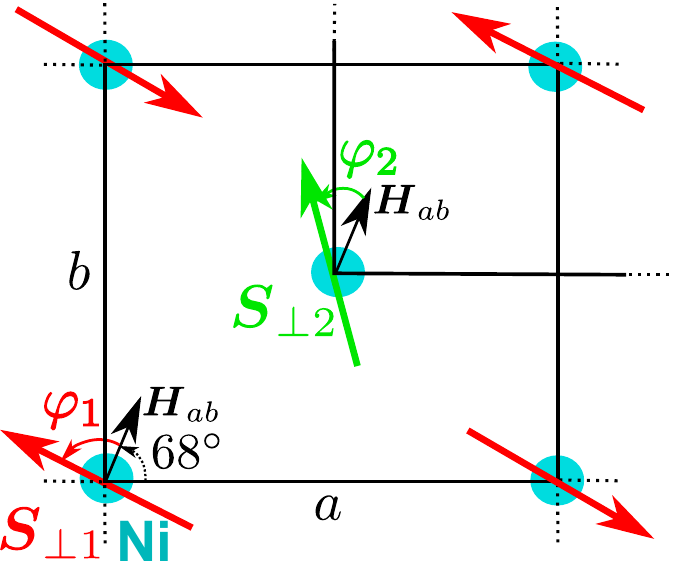}
\caption{Sketch of a general magnetic structure (orientations of $\boldsymbol{S}_{\perp 1}$ and $\boldsymbol{S}_{\perp 2}$ vectors) that was anticipated in the fits. Crystallographic $ab$ plane corresponds to the plane of the figure, and $\boldsymbol{H}_{ab}$ is the projection of the field into this plane (its orientation, at $68^{\circ}$ from the $a$ axis, was determined from the fit of the nitrogen quadrupolar splittings - see the text). Red and green colors denotes the two different tetragonal subsystems present within the BCT lattice. For the final solution (A in Tab.~\ref{taborientations}), $\boldsymbol{S}_{\perp 1}$ and $\boldsymbol{S}_{\perp 2}$ are found to be mutually parallel and nearly perpendicular to $\boldsymbol{H}_{ab}$   ($\varphi_1$ = $\varphi_2$ = 96(4)$^\circ$). }
 \label{defphi1phi2fig}
\vspace{0.1cm}
\end{figure}

All $^{14}$N(2) line-splittings of the NMR spectra recorded at $H=12.22$~T and $11.27$~T
(see Fig.~1) 
were fitted globally, and three equivalent solutions were found for the order parameter,
having all the same \textit{amplitude} $S_{\perp}$ but differing in the
\textit{orientations} of the $\boldsymbol{S}_{\perp 1}, \boldsymbol{S}_{\perp 2}$ vectors (see Fig.~\ref{defphi1phi2fig} and Table~\ref{taborientations}).
Clear selection between these three solutions was obtained by comparing the three corresponding calculated predictions for the proton $^1$H(3) lines (providing the most reliable fits) to the observed proton spectra.\cite{Sblinderthesis}

In addition to the spectra shown in Fig.~1, a number of partial nitrogen spectra were acquired in order to track the magnetic field dependence of one convenient line-splitting among all $^{14}$N(2) lines, and one among $^{14}$N(1) lines. The absolute value of $S_\perp$ determined at $H=11.27$~T was then used to scale the field dependence of these two line-splittings, giving, after merging both sources of data, the order parameter amplitude curve shown in Fig.~2.

\begin{table}[h!]
\begin{tabular}{|c|c|c|}
\hline
    & $\varphi_1 (^{\circ})$        & $\varphi_2 (^\circ)$   \\  \hline
    \bf{solution A}         &$\bf{97 \pm 3}$    & $\bf{95 \pm 3}$   \\ \hline
    solution B & $36 \pm 3$ & $108 \pm 1$   \\ \hline
    solution C & $82 \pm 3$ & $41 \pm 3$    \\ \hline
\end{tabular}
\caption{Three solutions for the magnetic structure  ($\varphi_1, \varphi_2$ orientations described on
Fig.~\ref{defphi1phi2fig}) that provide equivalently good fits of $^{14}$N(2) splittings. Solution A was finally retained after comparing the $^1$H spectra to its simulation (see the text).}
\label{taborientations}
\end{table}

\pagebreak


\begin{thebibliography}{99}
\bibitem{noteAx}
   {The formulas in this Rapid Communication correspond to the expressions from Refs.~\onlinecite{bouillot2011} and
  \onlinecite{klanjsek2008} where we replaced the factor $A_x$  by $2 A_x$ as needed to switch from the spin-$1/2$ ladder to the spin-1 chain.}
\end{thebibliography}
\end{document}